\documentclass[%
 aip,
rsi,%
 amsmath,amssymb,
preprint,%
]{revtex4-1}

\usepackage{graphicx}
\usepackage{dcolumn}
\usepackage{bm}
\usepackage{color}

\begin{document}

\preprint{AIP/123-QED}

\title[Adhikari et al. 2019]{Reconnection from a Turbulence Perspective}

\author{S. Adhikari}
\email{subash@udel.edu.}

\affiliation{Department of Physics and Astronomy, University of Delaware, Newark, DE 19716, USA}

\author{M. A. Shay}%

\affiliation{Department of Physics and Astronomy, University of Delaware, Newark, DE 19716, USA}

\affiliation{Bartol Research Institute, Department of Physics and Astronomy, University of Delaware, Newark, DE 19716, USA}

\author{T. N. Parashar}%
\author{P. Sharma Pyakurel}%

\affiliation{Department of Physics and Astronomy, University of Delaware, Newark, DE 19716, USA}

\author{W. H. Matthaeus}%

\affiliation{Department of Physics and Astronomy, University of Delaware, Newark, DE 19716, USA}

\affiliation{Bartol Research Institute, Department of Physics and Astronomy, University of Delaware, Newark, DE 19716, USA}

\author{D. Godzieba}%

\affiliation{Department of Physics and Astronomy, University of Delaware, Newark, DE 19716, USA}

\author{J. E. Stawarz}%
\author{J. P. Eastwood}%

\affiliation{\mbox{Department of Physics, Imperial College London, SW7 2AZ, UK}}

\author{J. T. Dahlin}
\affiliation{\mbox{NASA Goddard Space Flight Center, Greenbelt, MD 20771, USA}}

\date{\today}

\begin{abstract}
The spectral properties associated with laminar, anti-parallel reconnection are examined using a 2.5D kinetic particle in cell (PIC) simulation. Both  the reconnection rate and the energy spectrum exhibit three distinct phases: an initiation phase where the reconnection rate grows, a quasi-steady phase, and a declining phase where both the reconnection rate and the energy spectrum decrease. During the steady phase, the energy spectrum exhibits approximately a double power law behavior, with a slope near -5/3 at wave numbers smaller than the inverse ion inertial length, and a slope steeper than -8/3 for larger wave numbers up to the inverse electron inertial length. This behavior is consistent with a Kolmogorov energy cascade and implies that laminar reconnection may fundamentally be an energy cascade process. Consistent with this idea is that the reconnection rate exhibits a rough correlation with the energy spectrum at wave numbers near the inverse ion inertial length. The 2D spectrum is strongly anisotropic with most energy associated with the wave vector direction normal to the current sheet. Reconnection acts to isotropize the energy spectrum, reducing the Shebalin angle from an initial value of 70 degrees to about 48 degrees (nearly isotropic) by the end of the simulation. 
The distribution of energy over scale is further analyzed by dividing the domain into 
spatial subregions and employing structure functions.

\end{abstract}

\maketitle

\section{Introduction}
\label{sec:level1}

Magnetic reconnection releases magnetic energy explosively and plays an important role in a wide range of plasmas, from laboratory to Heliospheric to astrophysical plasmas (e.g., \cite{Yamada10}). In the last decade, its role as an element of turbulence and a dissipation mechanism for turbulent energy has come under increasing scrutiny. Spacecraft observations have found that reconnection occurs associated with turbulence in Earth's magnetosheath~\cite{Retino07,Phan18} and is associated with coherent structures in the solar wind~\cite{Osman14}. Two-dimensional MHD, Hall MHD, and kinetic PIC simulations of turbulence have been used to study the statistics of reconnection, finding a large variation of reconnection rates at X-lines~\cite{servidio2009magnetic,servidio2010statistics,Haggerty17}. This reconnection can heat the plasma and energize particles~(e.g., \cite{Karimabadi13},\cite{Shay18}). In some observations, the reconnection does not couple to the ions~\cite{wilder2017multipoint,Phan18} because the length scales associated with the turbulence are too small~\cite{sharma2019transition,stawarz2019properties}, a fact born out by simulations of turbulence~\cite{Haynes14,Califano18}. Finally, it has been suggested that reconnection in some cases modify the cascade of turbulent energy from large to small scales~\cite{Cerri17,Dong2018,Mallet17,Boldyrev17,Franci17,Papini18}.

On the other hand, the role of turbulence associated with reconnection has received significant attention. Reconnection is known to generate several secondary instabilities~(e.g., \cite{strauss1988turbulent,Buechner97,Drake97,Loureiro2007} and references therein) which themselves have been shown to generate energy spectra with power laws consistent with a turbulent cascade~\cite{pucci2018generation,munoz2018kinetic,lu2019turbulence}. Turbulent fluctuations in the plasma flowing into the reconnection region have been found to affect the process of reconnection~\cite{matthaeus1986turbulent}, and even amplify the reconnection rate~\cite{lazarian1999reconnection,matthaeus1985rapid}.

There remains a fundamental unanswered question, however, related to the interplay of turbulence and reconnection.  Namely, what does magnetic reconnection look like from a turbulence perspective, even in a 2D configuration ordinarily considered to be laminar? In addressing this, it is important to be clear that we include the individual components -- the inflow region, the diffusion, exhaust, separatrix regions (which we will group together), and the reconnected magnetic islands-- as elements of the reconnection process as a whole. We group the regions in this way because the inflow region supplies the energy for reconnection, the diffusion/separatrix/exhaust regions is where much of the dynamical activity occurs, and the island regions are a major consequence of the reconnection process. (Later we will call these regions III, II and I, respectively.) We will make some attempts at diagnosing their separate properties but ultimately we must treat these as strongly interacting components of a nonlinear system as we seek to understand the general relationship between reconnection and properties often associated with turbulence. As an example, what are the spectral properties of laminar reconnection and can they be related to various characteristics of reconnection (such as the reconnection rate)?

A typical example of laminar reconnection is seen in the GEM Challenge studies~\cite{Birn01}. Such an understanding of the spectral properties of laminar reconnection could act as a baseline, allowing more accurate determination of the role of reconnection in turbulence and vice versa. Additionally, it may also shed light on the fundamental properties of reconnection. Such a connection between turbulence and reconnection has been suggested previously based on modest 2D MHD simulations~\cite{matthaeus1986turbulent}.

 To address these questions we study the spectral properties of laminar, anti-parallel reconnection using 2.5D kinetic particle-in-cell (PIC) simulations. No turbulent fluctuations are manually added to the system, although random noise associated with the finite number of particles-per-cell is present. Strikingly, the laminar reconnection process generates a magnetic spectral density with a power law near $-5/3$ for $k \,d_i \lesssim 1$ and a steeper power law for larger $k$, where $d_i \equiv c/\omega_{pi}$. We emphasize that this power-law behavior is occurring in a 2D simulation where the reconnection current layers and separatrices do not exhibit the fluctuations often seen in 3D reconnection simulations~(e.g.~\cite{Buechner97}). To better understand the distribution of fluctuations in real space we will
 employ two-point structure functions \cite{biskamp2003magnetohydrodynamic} to resolve contributions from 
 distinct regions. 
 This overall behavior reinforces the interesting possibility that even laminar reconnection in a kinetic plasma fundamentally involves an energy cascade process\cite{matthaeus1986turbulent}. 

The paper is organized as follows. In section \ref{sec:level2} we introduce the details of the simulation and list all the parameters assigned. Section \ref{results} discusses the results and findings of the study. Finally, in section \ref{conclusion} we present the conclusions and discussions of the research.

\section{Simulation}
\label{sec:level2}

Laminar reconnection is studied by performing kinetic simulations using the particle in cell (PIC) p3d code \cite{zeiler2002three}. The simulation is of anti-parallel reconnection (no guide field) carried out in 2.5 dimensions. In this simulation, magnetic field is normalized to $B_{0}$ and the number density is normalized to a reference density $n_{0}$. All the lengths are normalized to ion inertial length $d_{i}=c/w_{pi}$, where $w_{pi}=\sqrt{4\pi n_{0} e^2/m_{i}}$ is the ion plasma freqency, time is normalized to the ion cyclotron time ($w_{ci}^{-1}=(eB_{0}/m_{i}c)^{-1}$), speed is normalized to the ion Alfv\'en speed ($v_{A0}=\sqrt{B_{0}^2/4\pi m_{i}n_{0}}$). Electric field is normalized to $E_{0}=v_{A0} B_{0}/c$ and temperature is normalized to $T_{0}=m_{i}v_{A0}^{2}$. The simulations have been performed in a periodic square box of various lengths $L_{x}=L_{y}=[51.2 d_{i},102.4 d_{i},204.8 d_{i}]$ all with a grid spacing of $\Delta =0.05 d_{i}$, and a time step of $\Delta tw_{ci}= 0.01$. Among the other parameters, the speed of light $c=15$, the electron and ion temperature are initially set as $T_{e}=0.25$, $T_{i}=1.25$, the mass ratio is $m_{i}/m_{e}=25$, the background density is $0.2$ and the half width of the current sheet $w_{0}$ is varied from $1.5 d_{i}$-$3d_{i}$ based on the size of the box. The plasma parameter for ions $\beta_{i}=2n_{i}T_{i}/B^{2}=0.5$ and that for electrons $\beta_{e}=2n_{e}T_{e}/B^{2}=0.1$. All of these simulations are performed in absence of a guide field $B_{g}=0$ and evolved until there is no more reconnection. This paper presents the results from the largest simulation $204.8d_{i}\times 204.8d_{i}$, where the half width of the current sheet is $3d_{i}$. 

The coordinate axes is chosen such that $\hat{x}$ represents the outflow, $\hat{y}$ represents the inflow and $\hat{z}$ represents the out of plane direction. The system is initialized with a double Harris current sheet equilibrium whose magnetic field is given by:
\begin{equation}
    \textbf{B} = 
 \left[\tanh \left ( \frac{y-L_{y}/4}{w_{0}}\right )-\tanh \left (\frac{y-3L_{y}/4}{w_{0}}\right )+1\right]\hat{x}.
\end{equation}
$T_i$ and $T_e$ are initially spatially uniform and the density $n$ is 0.2 outside the current sheets and varies to maintain total pressure balance. A perturbation of the form,
\begin{equation}
    \mathbf{\widetilde B}=\psi_{0} \frac{2\pi}{L_{x}} \cos\left(\frac{2\pi x}{L_{x}}\right)\hat{y}
\end{equation}
is applied to initiate reconnection, where $\psi_{0}$ is the initial strength of the perturbation chosen as $\psi_{0}=0.12$. This perturbation introduces two X-lines at ($L_{x}/4$,$3L_{y}/4$) and ($3L_{x}/4$,$L_{y}/4$). The strength of perturbation, $\psi_{0}$ is chosen such that the initial width of the magnetic island is about the half width of the current sheet.

\section{Results}
\label{results} 

\begin{figure}
 
\includegraphics[width=3.3in]{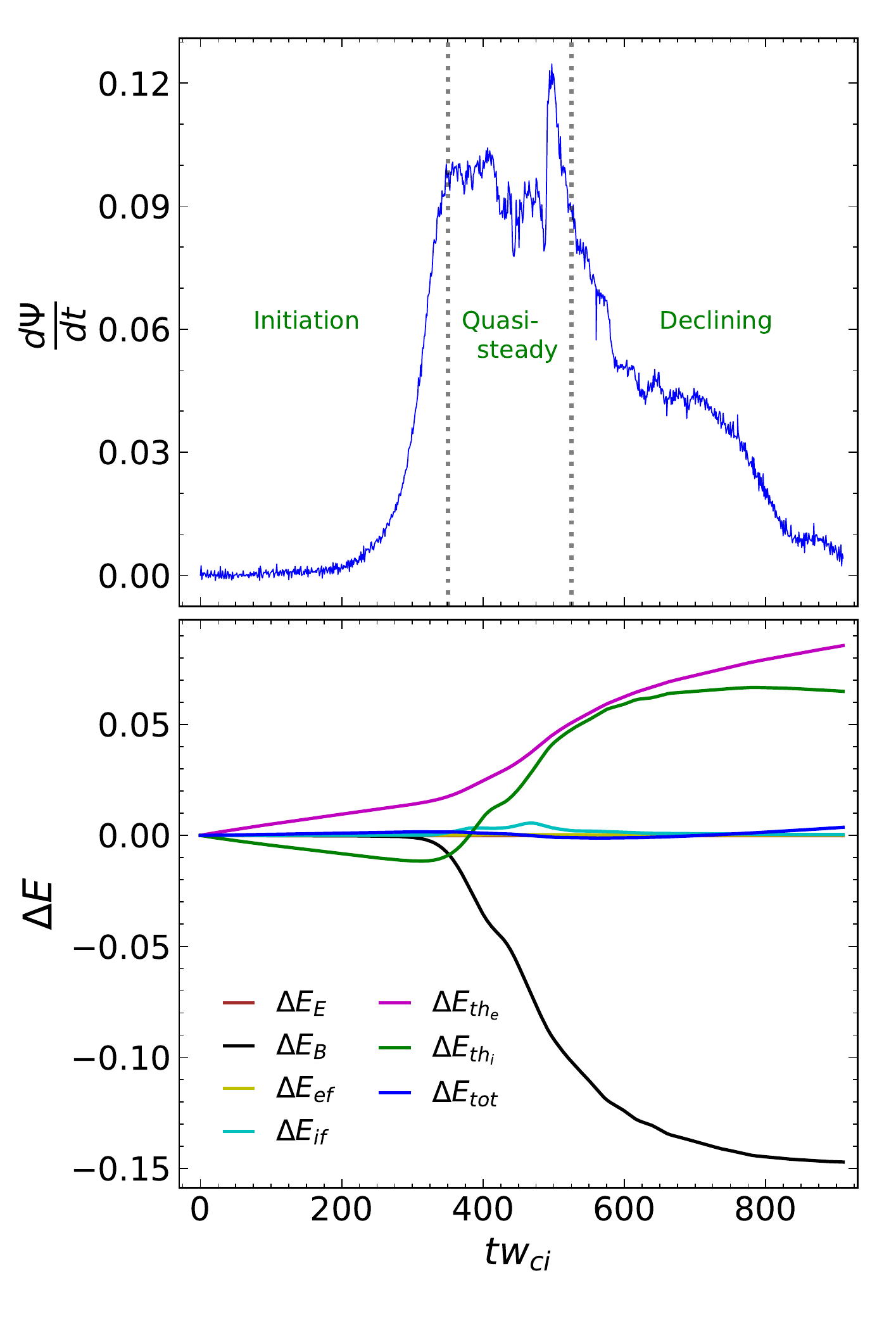}
\caption{ Reconnection rate of the simulation (top), and change in various forms of energy over time (bottom). The evolution of reconnection in the system is divided into three phases, shown in green (top panel). The energy curves plotted are the change in electrical energy ($\Delta E_{E}$), magnetic energy ($\Delta E_{B}$), thermal energy of ions and electrons ($\Delta E_{th_{i,e}}$), flow energy of ions and electrons ($\Delta E_{f_{i,e}}$), and the total energy ($\Delta E_{tot}$) which is the sum of all the above mentioned energies. }\label{fig:rrate}
\end{figure}

Fig.~\ref{fig:rrate} shows the rate of  magnetic reconnection in the upper X-line of the simulation (top panel), and the change in different forms of energy in the system (bottom panel).  The reconnection rate is calculated as the rate of change of magnetic flux ($\Psi$) between the X- and O- line. The simulation evolution can roughly be broken into three time periods as denoted in Fig.~\ref{fig:rrate}. During the initiation phase, the reconnection rate slowly increases and then suddenly  accelerates around $t \approx 250$. During the quasi-steady phase from  $t \approx 350$ to
$t \approx 525$, the reconnection rate is relatively steady with a value of $0.1$. The magnetic energy steadily decreases during this phase with the released energy going primarily into electron and ion thermal energy. A minimal amount of energy transfers to ion bulk flow energy, and total energy in the system is conserved quite well. The peak in the reconnection rate near $t \approx 500$ is associated with the formation of a secondary magnetic island. During the declining phase, the reconnection rate drops, plateaus at a rate near 0.05, and ultimately drops nearly to zero. During this phase the rate of magnetic energy change in Fig.~\ref{fig:rrate} gradually reduces. 

\begin{figure}
\hspace{0cm}
\includegraphics[width=3.2in]{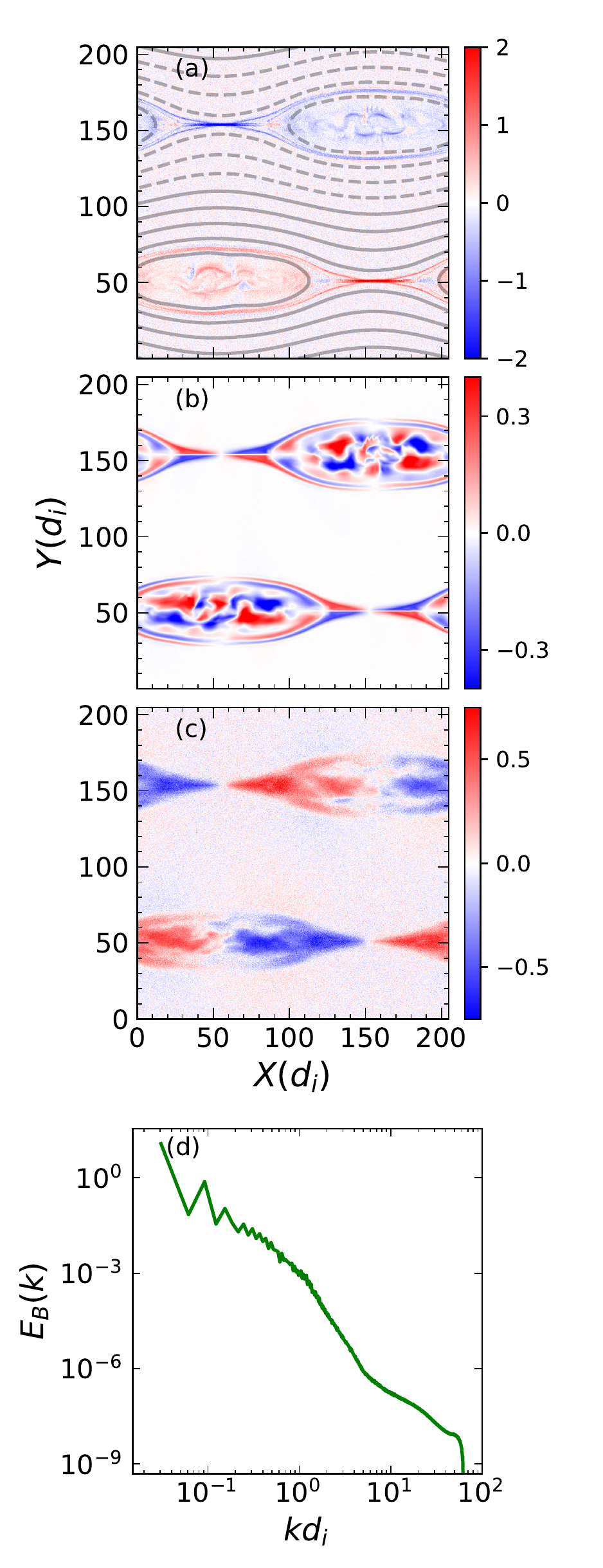}
\caption{General features of steady laminar reconnection at $t = 467.6$: (a) out of plane electron velocity ($v_{ez}$) with magnetic field lines (lines of constant magnetic vector potential along $\hat{z}$), (b) out-of-plane magnetic field ($B_{z}$), (c) ion outflow velocity ($v_{ix}$), and (d) omni-directional magnetic energy spectrum.}\label{fig:vix} 
\end{figure}
Fig.~\ref{fig:vix} shows an overview of the reconnection geometry during the steady phase at $t = 467.6$: (a) out of plane electron velocity ($v_{ez}$) with contours of the $z-$component of the magnetic vector potential, (b) the out of plane magnetic field ($B_{z}$), (c) the outflow velocity of ions ($v_{ix}$), and (d)the omni-directional magnetic energy spectrum~\cite{matthaeus2007spectral}. The magnetic field lines show the formation of a system size magnetic island as a result of breaking and reconnection of the field lines. In the vicinity of the X-lines, there is a dynamically appearing quadrupolar perturbation to $B_{z}$ associated with Hall physics~\cite{Mandt94}. Inside the magnetic island in a region known to become turbulent \cite{pucci2018generation}, $B_z$ has a complicated structure due to the complex electron flows.

The omni-directional magnetic spectrum in Fig.~\ref{fig:vix}d is calculated from the total magnetic field throughout the entire simulation domain. During this steady reconnection period, the omni-directional magnetic spectrum surprisingly acts roughly as a double power-law for $k\,d_i < 5$, with a break near $k\,d_i \sim 1$. This double power-law is strikingly similar to both kinetic simulations~\cite{parashar2009kinetic,howes2008kinetic} and observations~\cite{alexandrova2009universality,leamon1998observational,eastwood2009observations} of turbulence.

\begin{figure*}
\vspace{0cm}
\hspace{-1cm}
\includegraphics[width=6.0in]{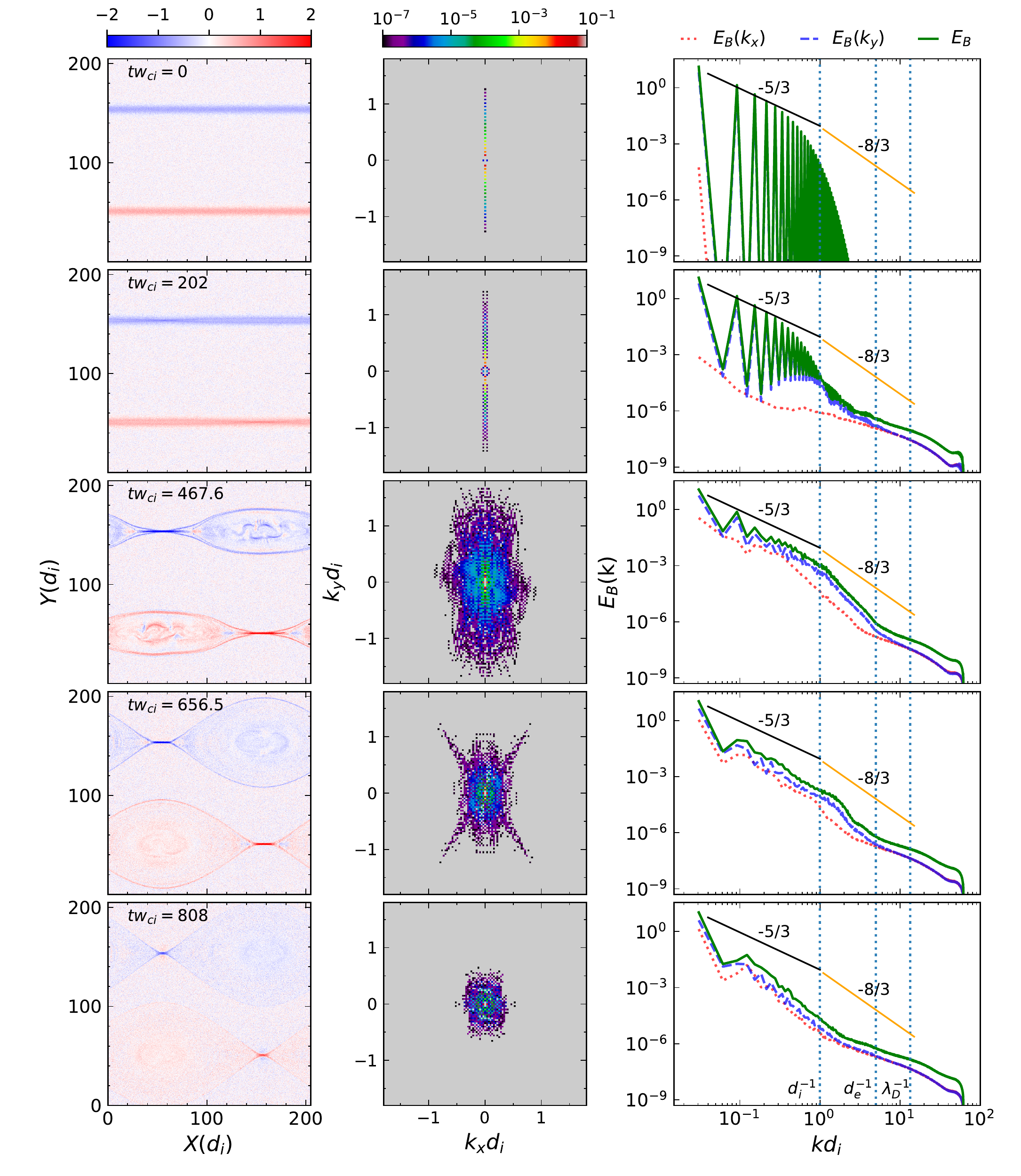}
\caption{Behavior in time of reconnection magnetic spectra features with (left)~$v_{ez}$, (middle)~2D energy spectrum, and (right) 1D reduce spectra with $E_B(k_x)$ in red, $E_B(k_y)$ in blue, and omnidirectional $E_B$ in green. Each row corresponds to the same time, given on the top left of $v_{ez}$ plot. The dotted vertical lines in the 1D spectra on right represent k values corresponding to $d_{i}^{-1}, d_{e}^{-1}$ and $\lambda_{D}^{-1}$. Solid lines with given slopes are drawn for reference; note that these slopes are denoted in the same location in the same plot, allowing estimation of change in magnitude of the spectra.}\label{fig:all} 
\end{figure*}
To better understand the relationship between laminar reconnection and its associated magnetic spectra, we now explore the evolution of the spectra as the system evolves. Fig. \ref{fig:all} shows the profile of out of plane electron velocity ($v_{ez}$) (left panel), the two-dimensional magnetic energy spectrum in the k-space (middle panel), and the omni-directional and reduced spectra of magnetic energy (right panel). Energy values smaller than $10^{-7}$ are omitted in the 2D spectrum. The wavenumbers corresponding to the ion inertial length ($d_{i}$), electron inertial length ($d_{e}$) and the Debye length ($\lambda_{D}$) are represented by the dotted vertical lines. The solid black and blue lines drawn for reference have the spectral index of $-5/3$ and $-8/3$.

At $t=0$, the initial double Harris current sheet condition is evident. The spectrum is strongly anisotropic, with almost all energy concentrated at $k_x = 0$ in the 2D spectrum. The 1D spectra show sharp features associated with the initial condition. At $t=202$, the reconnection rate is still quite small and there is a slight narrowing of $v_{ez}$ in the vicinity of the X-lines and a slight broadening near the O-lines. In the 2D spectrum a slight broadening of the spectrum along $k_x$ has occurred near $|k\,d_i| \sim 1.$ In the 1D spectra, a large increase in the energy spectra for $k\,d_i \gtrsim 1$ is evident. Much of this increase is due to the impact of noise due to finite particles per grid on the magnetic field, especially at the highest values of $k.$ 

At $t=467.6$ in Fig.~\ref{fig:all} the reconnection is quasi-steady, as observed in Fig. \ref{fig:rrate}. The global magnetic islands have grown considerably, reaching an island width of about 20. Associated with this magnetic island growth is a significant isotropization of the 2D magnetic spectrum, which has broadened considerably in the $k_x$ direction. This isotropization is similar to what was seen in 2D MHD reconnection initialized with broadband background turbulence~\cite{matthaeus1986turbulent}. During the quasi-steady reconnection period, for $k\,d_e < 1$ the omnidirectional power spectrum is roughly a double powerlaw. The omni-directional spectrum exhibits a power law of spectral index $-5/3$ in the inertial range $(k\,d_i < 1)$, characteristic of Kolmogorov spectrum \cite{kolmogorov1991local,kraichnan1967inertial} before steepening above $d_{i}^{-1}$. In between the ion and electron inertial range, $d_{i}^{-1}<k<d_{e}^{-1}$ the spectrum is reasonably well represented by a power index of $-11/3$.  This range of spectral slopes is in agreement with the experimental results of solar wind turbulence ~\cite{leamon1998observational,smith2006dependence,alexandrova2008small,sahraoui2009evidence}. Note also that simulations with both smaller and larger simulation domains (not shown) display similar spectral features.

At $t=656.5$ in Fig.~\ref{fig:all} the evolution has entered the declining phase, where the reconnection rate decreases. The magnetic island widths have roughly doubled in size, and the 2D spectrum has further evolved towards isotropy. A notable feature in the 2D spectrum are wings extending into all four quadrants, roughly making an angle of about $55^{\circ}$ with the horizontal. This angle is associated with the angle of the separatrices in the left column. In the 1D spectra, the slope in the inertial range $(kd_{i}<1)$ has steepened considerably. 
It is noteworthy also 
that the steepening of the spectra to a slope greater than -8/3 now occurs at a larger $k$ $(kd_{i}>1)$. 

Generally speaking, 
from a plasma turbulence perspective 
the spectral steepening 
and movement of the breakpoint near $kd_i=1$ 
that we see here in Fig. \ref{fig:all}
is quite interesting.
In this regard it is noteworthy that the behavior of the breakpoint 
between MHD-like scales (for $\beta<1$,  $kd_i <<1$) 
and subproton kinetic scales ($kd_i\gtrsim 1$) is a widely studied topic in studies of 
strong plasma turbulence, including the solar wind \cite{leamon1998observational,ChenEA14-grl,Alexandrova13}
and magnetosheath \cite{ChhiberEA18}, as well as simulations \cite{HowesEA08,Karimabadi13,Franci17}. 
Given the documented complexity of this issue we will not attempt a full explanation 
of the kinetic scale evolution seen in spectra and other diagnostics here.
However it seems certain that the processes observed at $kd_i>1$ are not fully determined by the 
MHD cascade but rather involve nonlinear kinetic 
plasma physics processes. 

Finally, at $t=808$ in Fig.~\ref{fig:all}, the electron current sheet has reduced considerably at the X-line. The 2D spectrum is almost completely isotropic, and the 1D spectra have steepened considerably for $k\,d_i < 2$.

\begin{figure}
\includegraphics[width=3.3in]{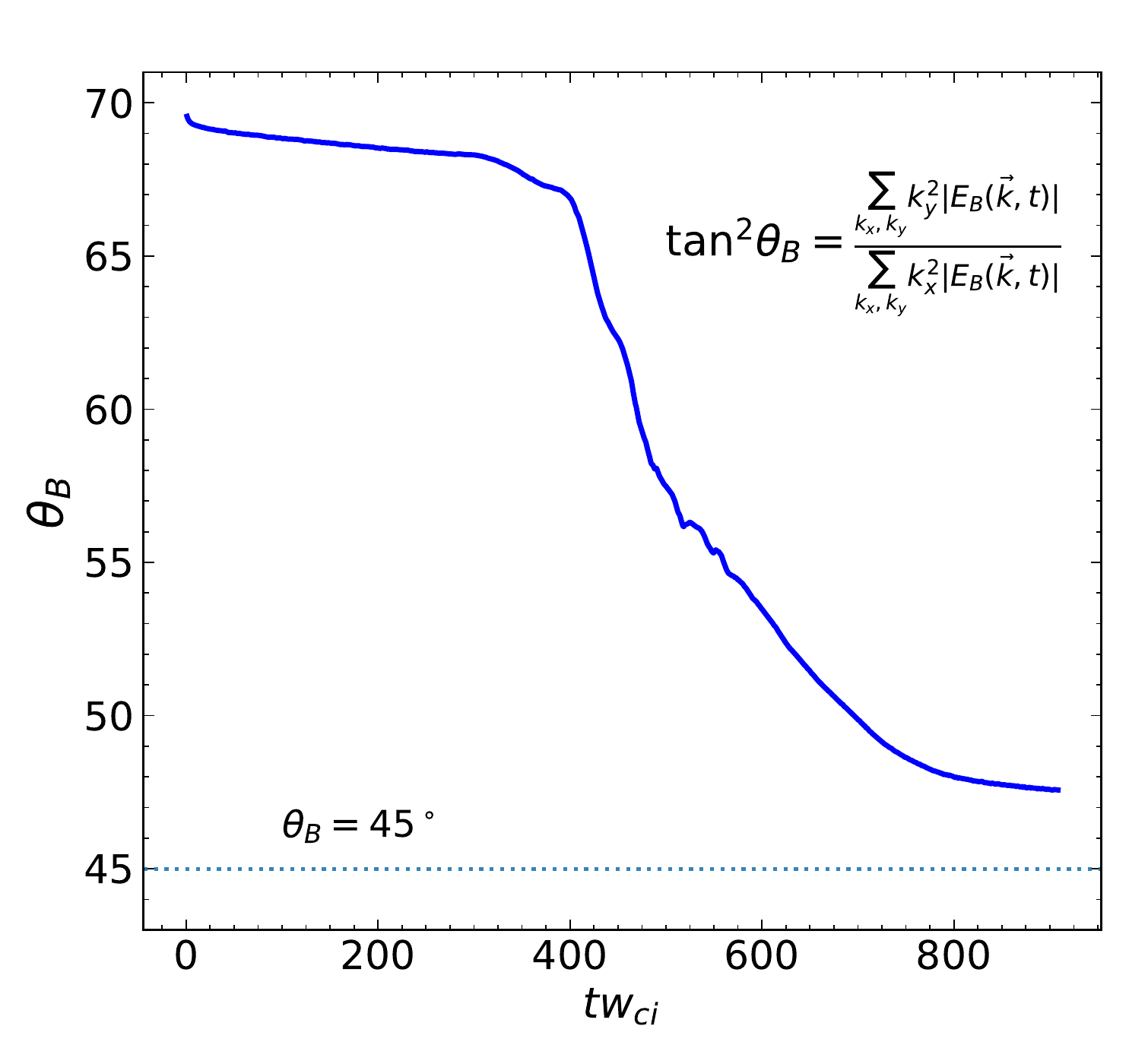}
\caption{Evolution of Shebalin angle, defined on the upper right. $E_{B}(\vec{k},t)$ is the two-dimensional magnetic energy. The dotted line at $\theta_{B}=45^{\circ}$ represents perfect isotropy.  }\label{fig:shebalin} 
\end{figure}
The evolution of the spectral anisotropy is represented by the Shebalin angle~\cite{shebalin1983anisotropy}. The Shebalin angle corresponding to the magnetic spectrum is defined as:
\begin{equation}
    tan^2\theta_{B} =\frac{\sum \limits_{k_{x},k_{y}} k_{y}^2|E_{B}(\vec{k},t)|}{\sum \limits_{k_{x},k_{y}} k_{x}^2|E_{B}(\vec{k},t)|},
\end{equation}
where $k_{x}$ and $k_{y}$ correspond to the wavenumber along $x$ and $y$ axes respectively while $E_{B}$ is the 2D magnetic energy.
Note that $\theta_{B}=45^\circ$ for an isotropic 2D spectrum.  Fig. \ref{fig:shebalin} shows the evolution of Shebalin angle in time. Initially the Shebalin angle is about $70^\circ,$ which may seem counterintuitive since the initial magnetic field has little variation along $x$. However, random particle fluctuations associated with the finite particles per cell create isotropic magnetic fluctuations which reduce the angle. The angle is nearly constant  for the first half of the simulation and then suddenly begins to decrease soon after the onset of quasi-steady reconnection. 

Interestingly, in Fig.~\ref{fig:shebalin}, there is a time lag of about $50$ between the onset of steady reconnection and this sharp decrease in the Shebalin angle. 
This time lag is comparable to the global nonlinear time, or eddy turnover time
of the system. 
A scale-dependent nonlinear time 
is defined as $\tau_{nl} = 1/(k\,u_k)$ with $u_k$ a 
characteristic velocity for a given $k$ value. 
For the global nonlinear time, 
we take $k = 2\pi/L = 2\pi/(204.8) \approx 0.03$ and $u_k$ as the Alfv\'en speed outside the current sheets, 
giving a global nonlinear time of $1/(0.03 \cdot 1) \approx 30$. 
Therefore, one possible explanation for the time lag is that a cascade of energy to 
smaller scales begins during the fast, quasi-steady phase, and it takes roughly 
one nonlinear time for the energy to cascade to the smallest 
scales and isotropization to begin. For $t \gtrsim 400$, the Shebalin angle steadily decreases, 
reaching almost full isotropy $(\theta_B = 45^\circ)$ by the end of the simulation. 
In fact, as we shall presently see, the initiation of fast reconnection releases 
magnetic energy that quite rapidly ($\sim$~one nonlinear time) unleashes a chain of dynamical effects that increasingly resemble turbulence. 

\begin{figure}
\includegraphics[width=6.7in]{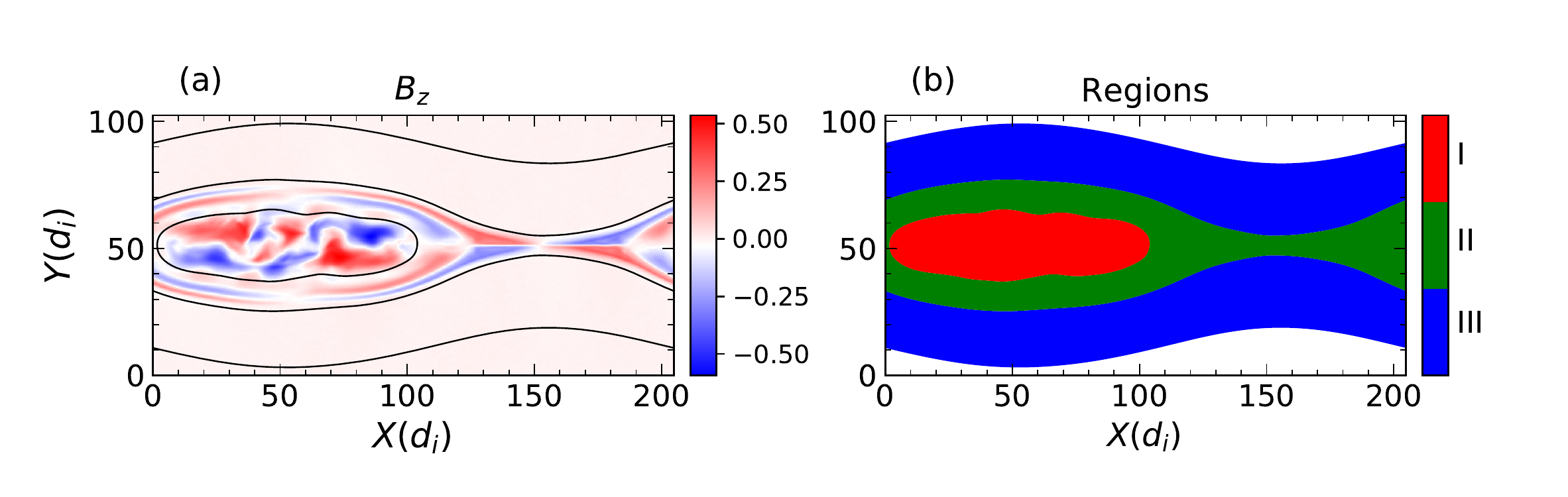}
\caption{Lower half of the simulation at $t = 467.6$.  (a) Out-of-plane magnetic field $(B_{z})$ and magnetic field lines used to separate spatial regions for analysis. (b) Region I (red) is the magnetic island, Region II (green) is the diffusion, exhaust, and the separatrix (DES region), and Region III (blue) is the inflow region.}\label{regions} 
\end{figure}

Given that the system we are studying is highly inhomogeneous and ansisotropic, 
questions naturally arise concerning the contributions to the spectrum from different spatial regions. 
Key spatial regions in the simulation are (I) the magnetic islands, (II) the diffusion region, exhaust, and the separatrix (collectively called DES), and (III) the inflow regions. 
The boundaries of those regions are illustrated 
in Fig.~\ref{regions} at $t = 467.6$.
The regions are bounded 
by appropriately chosen magnetic field lines and are highlighted with different colors in Fig.~\ref{regions}b. 
Practical difficulties arise in calculating spectra for such regions using standard methods. 
For example, familiar methods based on Fast Fourier Transforms are global in nature 
and are not easily adapted to complex regions.

As an alternative to spectra, 
here we compute the second-order structure functions in each region following the methods of Pucci et al., (2018) ~\cite{pucci2018generation}. 
The structure function is defined as $S^{2}_{B}(\textbf{l})=<|\textbf{B}(\textbf{r}+\textbf{l)}-\textbf{B}(\textbf{r})|^{2}>$, 
where $\textbf{B}$ is the total magnetic field, $\textbf{l}$ is the lag in the real space, and the average ($< >$) is done over a given volume.
We note that the value of the structure function at a given spatial lag may be interpreted as the contribution to the 
energy from fluctuations at that scale. 
The upper and lower half of the box have almost identical structures, therefore the analysis of the upper box is not performed. 
The structure function is then computed in these three regions at $t=467.6$ when the reconnection is quasi-steady. 
The smallest lag is equal to $\Delta x =0.05$ while the maximum lag is chosen to be half of the size of the region.
For simplicity, we calculate the 
structure functions with one dimensional lags either along  $x\; (l_{x})$ or along $y\; (l_{y}).$ 

\begin{figure}
\includegraphics[width=6.7in]{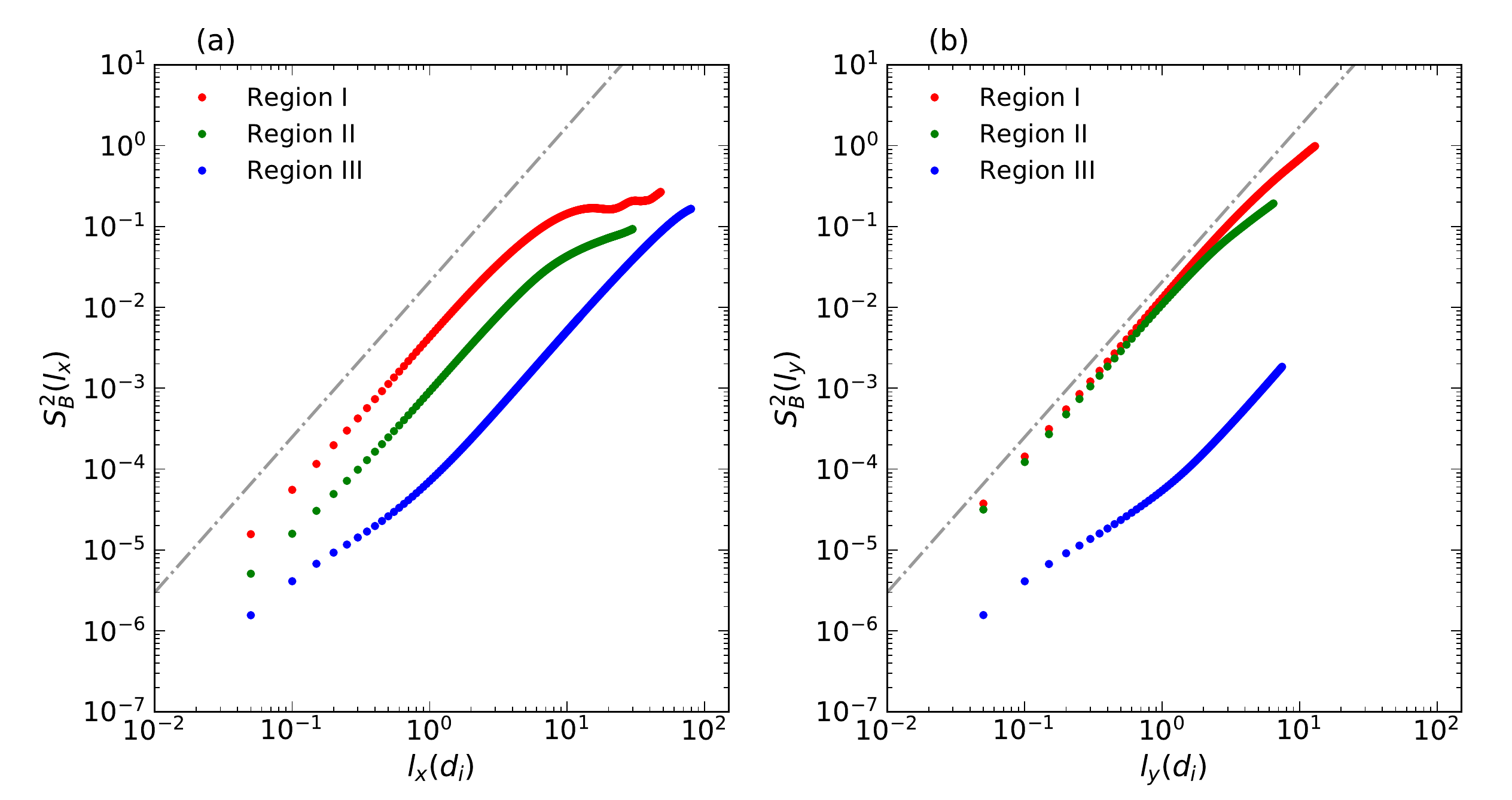}
\caption{Second order structure functions $S^{2}_{B}(l_{i})$ computed across the regions I (red), II (green), and III (blue) at $tw_{ci}=467.6$ with (a) lag along x, and (b) lag along y. The color index is the same as in Fig. \ref{regions}.
Identical reference lines with arbitrary slope are added
to facilitate comparison between the panels.}\label{stfxn} 
\end{figure}

In Fig.~\ref{stfxn} are shown the structure functions computed along 
$l_x$ and $l_y$, in separate panels. 
First, one observes that 
the inflow region (III) has the smallest structure function in both directions, indicating that the turbulence energy is 
lower in region III than the other regions at all scales shown. 
Second, anisotropy is present in both the island (I) and the DES (II) regions with $S^2(l_y) > S^2(l_x)$ 
for equal lags. However we note that the island region is closer to isotropic than is the DES region. To highlight this anisotropy, an identical reference line has been added for easy comparison of the two panels. It is clear that the structure function along $l_y$ makes the dominant contributions to the energy budget. The island (I) and the DES region (II) have nearly equal structure functions for $l_y \lesssim d_i$ and remain comparable in magnitude at all scales that can be compared. This implies that for these lags or length scales, both 
the island (I) and the DES region (II) 
play an important role in terms of contributions to the energy budget.

\begin{figure}
\includegraphics[width=3.5in]{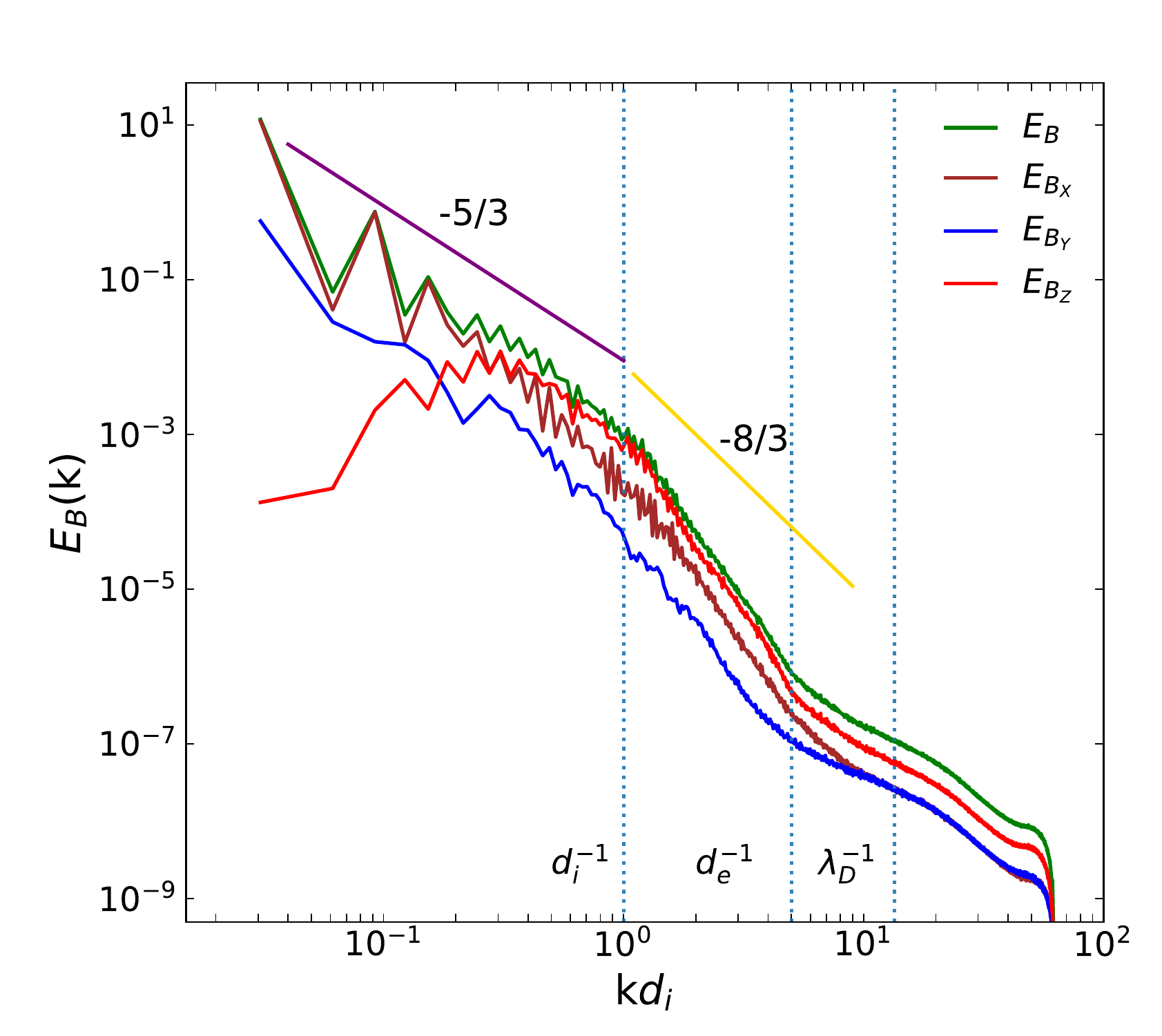}
\caption{Contribution of each component of the magnetic
field to the omni-directional spectrum along with the total omni-directional spectrum
at $tw_{ci}=467.6$.}\label{fig:omnii} 
\end{figure}
 Fig. \ref{fig:omnii} shows the omni-directional spectra of the three components of the magnetic field and also the total magnetic field when the reconnection is quasi-steady. For $k\,d_i \lesssim \frac{1}{2},$ the total omni-directional spectrum is primarily due to the x-component of the magnetic field $B_x$, which is the reconnecting field. However, at smaller length scales ($k\,d_i \gtrsim 1$) the contribution is dominated by the out of plane magnetic field ($B_z$). This transition is consistent with the onset of Hall physics for $k\,d_i \sim 1$. Curiously, the approximate $-5/3$ slope extends to $k$ values larger than the transition between $B_x$ and $B_z$ dominance; this behavior is consistent with the generation of $B_z$ at $k\,d_i > 1$ and then the transfer of the $B_z$ structure to longer wavelengths. An example of this is seen in Fig.~\ref{fig:vix}c at $x \approx 180$ for the lower x-line where the width along $y$ of $B_z$ is nearly 20, which corresponds to $k \approx 1/3$. This type of large scale $B_z$ in the reconnection exhaust has also been seen in observations of reconnection in the Earth's magnetosheath~\cite{Phan07}.

 \begin{figure}
\includegraphics[width=3.3in]{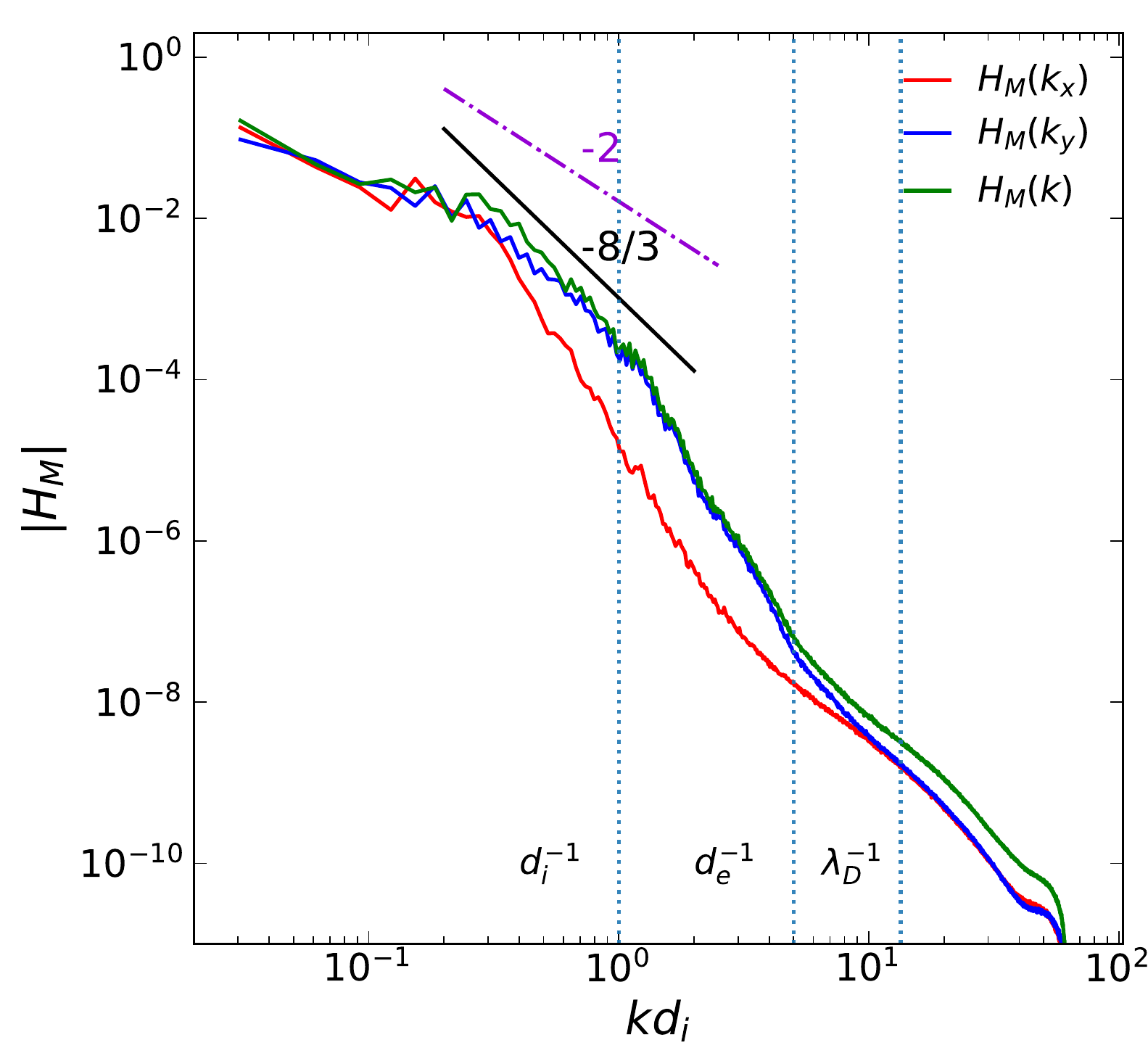}
\caption{Reduced (red, blue) and omni-directional (green) spectra of magnetic helicity at $tw_{ci}=467.6$. A dash-dotted line with slope $-2$ (violet) and a solid line with slope $-8/3$ (black) is drawn for reference.}\label{helspec} 
\end{figure} 
In order to further explore the possibility of the cascade properties of this system, we examine the spectrum of the magnetic helicity. Magnetic helicity is defined as $H_{M}= \int \textbf{A} \cdot \textbf{B}\, dV$, where $\textbf{B}$ is the magnetic field, $\textbf{A}$ is the magnetic vector potential ($\textbf{B}=\nabla \times \textbf{A}$), and $V$ represents the entire volume of the simulation. In the Fourier space, magnetic helicity is computed as  $H_{M}= {\Re}\left\{ \;\sum_{k} \textbf{a}_{k}\cdot \textbf{b}^{*}_{k}\; \right\}$, where $\textbf{a}_{k}$ and $\textbf{b}_{k}$ are the fourier transform of the vector potential and the magnetic field respectively and $\textbf{b}^{*}_{k}$ is the complex conjugate of $\textbf{b}_{k}$ \cite{biskamp2003magnetohydrodynamic}. Fig.~\ref{helspec} shows the magnitude of the reduced ( $H_M (k_x)$ and $H_M (k_y)$ ) and omni-directional ( $H_M (k)$  ) magnetic helicity in k-space when the reconnection is quasi-steady, $tw_{ci}=467.6$. The helicity is dominated by the $z-$components of the vector potential and magnetic field. Using dimensional arguments it can be shown that the omni-directional magnetic helicity spectrum has a power law slope of -2 for the inverse cascade scenario and -8/3 for a constant fractional helicity $( \sigma_m = k\,H_M / E_B )$ and a Kolmogorov magnetic spectrum~\cite{frisch1975possibility,biskamp2003magnetohydrodynamic}. For $k\,d_i \lesssim 1,$ the slope is roughly consistent with the constant fractional helicity case. 
 
\begin{figure}
\includegraphics[width=3.4in]{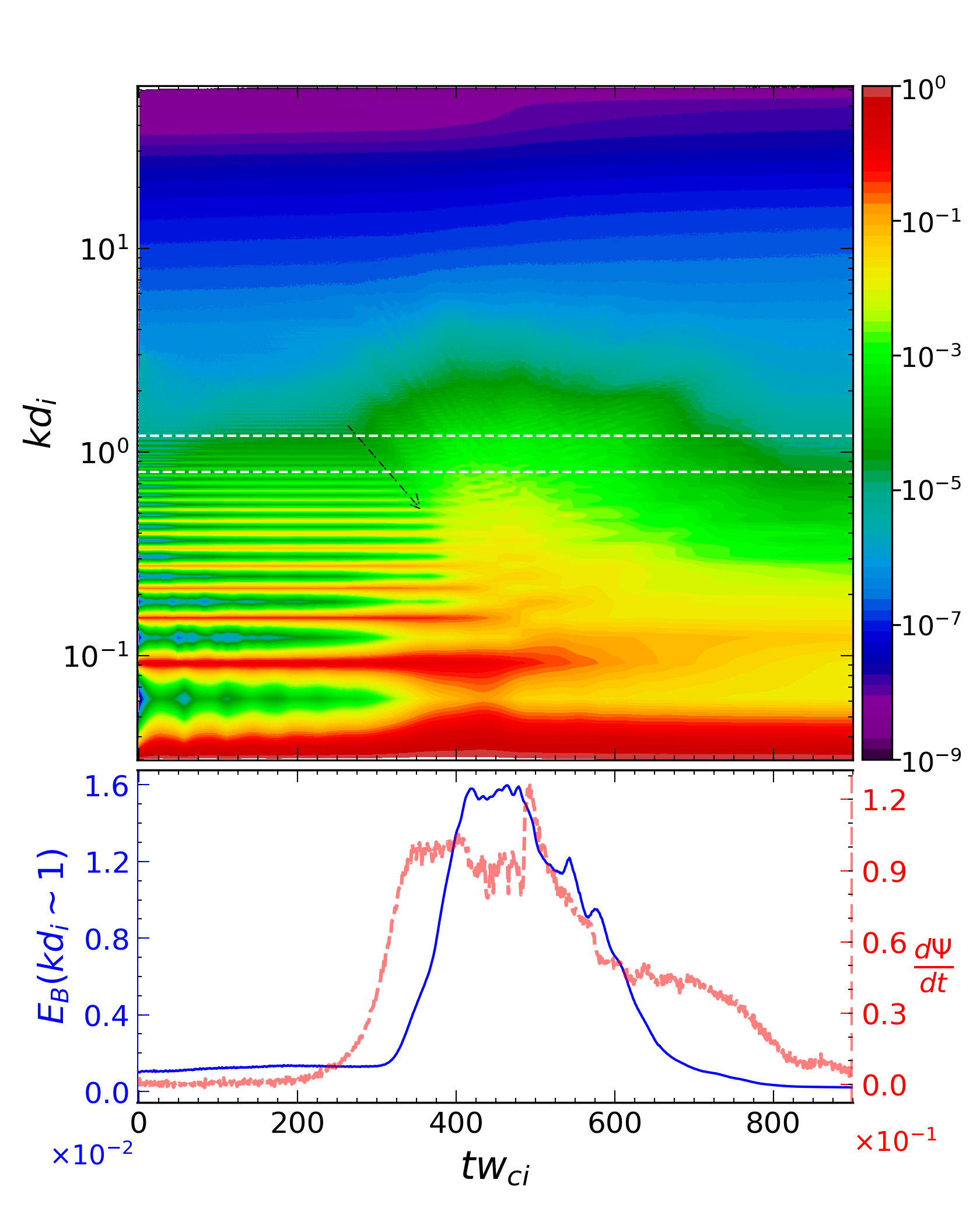}
\caption{Spectrogram of the omni-directional magnetic
energy spectrum (top) and the time evolution of the magnetic energy at $kd_{i}\sim
1$ (blue) along with the reconnection rate (red) (bottom). The dotted white lines on
the upper plot are drawn at $kd_{i} = 0.8$ and $kd_{i} = 1.2$. The black arrows show the migration of the change in the spectrogram (See text for details). }\label{fancy} 
\end{figure} 
Further insight into the spectral properties of the reconnection in this system can be gained by examining a spectrogram of the omnidirectional magnetic spectrum, shown in Fig.~\ref{fancy} (top panel). At early time, the spectrum of the initial $\tanh$ Harris equilibrium is clearly evident. When steady reconnection initiates near $t \approx 350$, a sudden change in the spectrogram occurs at both the smallest $k$ values and also at high $k$ values. The sudden change in high $k$ appears to migrate 
towards lower values of $k$, and have drawn black arrows into Fig.~\ref{fancy} to highlight this behavior. 
At the same time, the broadening of several bands at 
low $k$ indicates transfer of energy towards the smaller scales. 
This is suggestive of bidirectional spectral transfer typical of turbulence 
with the net transfer being towards higher wavenumbers~\cite{verma2005energy}.  

The length scale $d_i$ plays an important role during antiparallel reconnection because it is roughly the width along the normal direction of the ion diffusion region, where MHD breaks down and the reconnection process begins. Intuitively, therefore, the magnetic energy spectra associated with $k\,d_i \approx 1$ may be intrinsically linked to reconnection properties. Indeed, dynamical activities such as current sheet intensification, formation of Hall fields, and even the formation of secondary islands would all be expected to influence the energy near $k\,d_i \sim 1$.   Therefore, in the spectrogram in Fig.~\ref{fancy}, we integrate the energy between the two dashed horizontal lines $ (0.8 \le k\,d_i \le 1.2)$ and plot the resultant energy as the blue curve in the bottom panel.  For reference the reconnection rate is shown in red. The correlation between the two is quite strong. First, the energy is quite steady during the period of steady reconnection, but with a time lag. Note that this time lag is comparable to the time lag noted for the Shebalin angle in Fig.~\ref{fig:shebalin}. The correlation with the drop in reconnection rate is quite striking.

\begin{figure}
\includegraphics[width=3.4in]{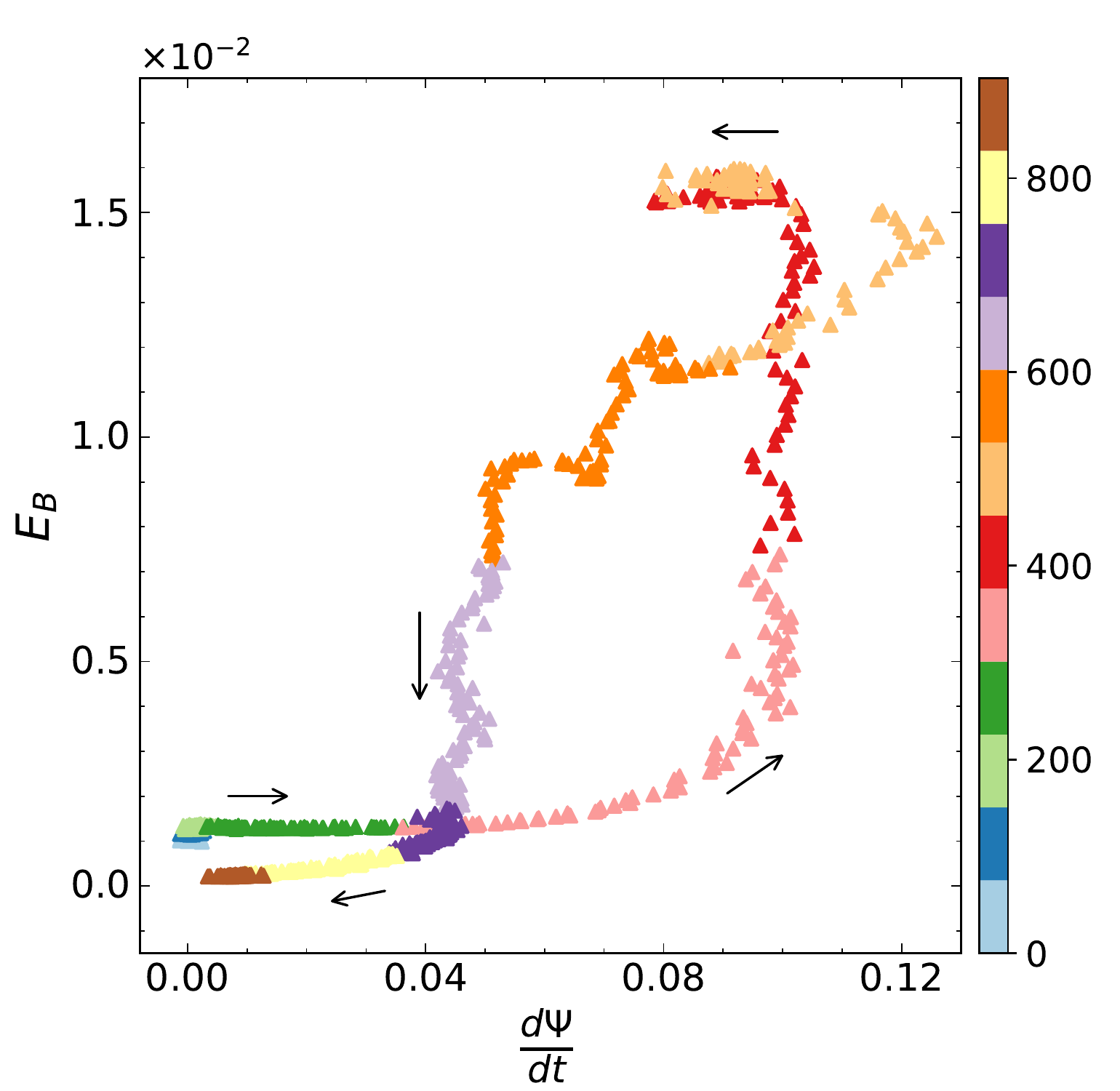}
\caption{ Scatter plot of the reconnection rate and the omni-directional
magnetic energy at k$d_{i}\approx 1$. The colorbar represents the time and
the arrow points in the direction of time. }\label{scatter}
\end{figure}
The connection between the reconnection rate and the magnetic energy at $kd_{i}\sim 1$ can also be visualized using a scatter plot as shown in Fig. \ref{scatter}. The behavior roughly resembles a hysteresis curve, indicating that the nature of the correlation changes during the evolution. Initially, the reconnection rate rises with little or no change in the energy. When the reconnection rate stabilizes, there is a sharp rise in the energy. The figure also reveals the drop in the magnetic energy during the late reconnection when the reconnection rate is almost constant at about $0.045$. The fall in the energy corresponds to a total time of $tw_{ci}\approx120$ ranging from $tw_{ci}=600-720$, which is about four times the nonlinear time ($\tau_{nl} = \frac{1}{k\,u_k}$) calculated at the largest scale of the simulation.  The time corresponds to the drop in the out of plane magnetic field inside the separatrix. The characteristic time for all of these processes appears to be comparable to the large scale $\tau_{nl}$ hinting at a connection with a turbulent energy cascade process. 

Finally, we 
call attention to the 
somewhat erratic behavior of the $d\psi/dt$-$E_B$ relationship seen in Fig. \ref{scatter}
at approximate times $500$ to $650$  in the declining phase of reconnection dynamics.
A similar epoch is observed in Fig.  \ref{fancy} and in subtle features of 
subproton spectral features in Fig. \ref{fig:all} in a similar time frame. 
We reiterate that these features are probably all related to complex plasma dynamics
that emerge here when the MHD driving is less strong \cite{Karimabadi13}. 
The basic physics of such behavior is actively discussed in space physics observations and in 
plasma simulation studies\cite{Alexandrova13,ChhiberEA18,ChenEA14-grl,HowesEA08}. 

\section{Conclusions}
\label{conclusion} 

In this paper we have examined the magnetic spectral behavior of laminar antiparallel reconnection using 2.5D fully kinetic PIC simulations. The omnidirectional magnetic energy spectrum and the rate of reconnection exhibit three discrete phases in time: (1) An initiation phase ( $t=0$ to $t \approx 350$ ) during which the reconnection rate and the energy spectrum grow. (2) A steady phase ( $t \approx 350$ to $t \approx 525$) where the reconnection rate stabilizes around 0.1. During this time, the energy spectrum exhibits approximately a double power law, with a slope near $-5/3$ for $k\,d_i \lesssim 1$ and a slope steeper than $-8/3$ for $d_i^{-1} \lesssim k \lesssim d_e^{-1}$. (3) A declining phase ( $t \approx 525$ to $t=900$ ) during which the reconnection rate gradually decreases, almost going to zero by the end of the simulation. During this phase, the spectrum gradually decreases for all k values and for $k\,d_i \lesssim 1$ the slope gradually becomes steeper than $-5/3$. 

We note that the double power law behavior during the steady phase is not simply an artifact of the initial spectrum: (1) The initial Tanh function itself actually generates a -2 spectrum at wavelengths long compared to the reciprocal current sheet width versus the -5/3 slope observed. (2) The Tanh generates only odd harmonics. However the spectrum that is generated dynamically fills in these “gaps”. (3) The approximate -5/3 spectrum extends to $k\,d_i \sim 1$ while the initial Tanh spectrum steepens well before that. And (4) the dynamically generated spectrum becomes progressively more isotropic, as described in the manuscript, whereas the initial Tanh spectrum is extremely anisotropic. 

The power-law behavior of the magnetic energy spectrum during the steady phase is consistent with the existence of a Kolmogorov energy cascade, which raises the intriguing possibility that laminar reconnection fundamentally involves an energy cascade process, even though the reconnection diffusion region and the separatrices do not exhibit strong fluctuations. However, proving that a cascade of energy is occurring will
require analysis of higher order cascade laws~\cite{politano1998karman,hellinger2018karman,bandyopadhyay2018incompressive} which will be discussed in a future
paper.

Initially, the magnetic energy is concentrated near very small $k_x$ with a range of $k_y$ consistent with the double $\tanh$ initial condition, leading to a Shebalin angle initially far from the isotropic value of $45^\circ$.  Consistent with previous MHD simulations~\cite{matthaeus1986turbulent}, the process of reconnection acts to isotropize the energy spectrum. During the steady phase, the Shebalin angle exhibits a sharp drop from $70^\circ$ to about $55^\circ$. During the declining phase, the angle reduces more slowly, ultimately reaching about $48^\circ$ at the end of the simulation.

As a further analysis of the distribution of energy in scale, 
we employed second order structure functions in three disjoint spatial
regions - the islands, the DES (diffusion, exhaust, and separatrix) region, and the inflow region.
We found that the inflow regions are quiescent as expected, 
while comparable contributions to the energy budget are due 
to the reconnection and island regions.

During the steady phase, the reconnecting magnetic field $B_x$ dominates the omnidirectional magnetic energy spectrum for $k\,d_i \lesssim 0.5$. For larger $k$, $B_z$ becomes the largest contributor. This behavior is consistent with Hall physics becoming important for $k\,d_i \gtrsim 1$. The ions are no longer frozen-in and the Hall term becomes important in Ohm's law~(e.g., \cite{Shay98a}). The reconnecting magnetic field is dragged out of the plane (along $z$)~\cite{Mandt94}, creating the quadrupolar Hall $B_z$ perturbation~\cite{Sonnerup79}. This Hall magnetic field is responsible for the observed magnetic helicity, but spectral analysis reveals that the helicity is not strong enough to drive an inverse cascade. Instead, the spectrum is closer to a constant fractional helicity spectrum that would be associated with a Kolmogorov direct energy cascade spectrum.

We find a correlation between the reconnection rate and the energy spectrum near $k\,d_i \sim 1$ (denoted as $E_{di}$), which makes sense as the ion diffusion region of reconnection has a width of a few $d_i$. Both the reconnection rate and $E_{di}$ rise during the initiation, reaching roughly constant values during the steady phase. Note, however, that there is a time lag, with $E_{di}$ plateauing about $50 w_{ci}^{-1}$ after the reconnection rate. This time lag is roughly comparable to the global nonlinear time (global eddy turnover time). If a cascade of energy is indeed occurring, a plausible explanation for the lag would be that it represents the time for energy to cascade from the energy containing scale to $k\,d_i \sim 1$.  The declining phase is marked by a sharp drop in both $E_{di}$ and the reconnection rate. 

In this manuscript we have found intriguing connections between laminar reconnection physics and turbulence phenomena expanding on previous suggestions based on MHD~\cite{matthaeus1986turbulent}. Future work will expand on this approach and may ultimately reveal the fundamental link between the two.

\begin{acknowledgments}
This research was supported by NASA Grants NNX17AI25G (Heliophysics Supporting Research)
and NNX14AC39G (MMS THeory and Modeling). 
We would like to acknowledge high-performance computing support from Cheyenne~\cite{Cheyenne18} provided by NCAR's Computational and Information Systems Laboratory, sponsored by the National Science Foundation. This research also used resources of the National Energy Research Scientific Computing Center (NERSC), a U.S. Department of Energy Office of Science User Facility operated under Contract No. DE-AC02-05CH11231.
\end{acknowledgments}

\bibliographystyle{abbrv}
\bibliography{laminar}

\begin{thebibliography}{10}

\bibitem{alexandrova2008small}
O.~Alexandrova, V.~Carbone, P.~Veltri, and L.~Sorriso-Valvo.
\newblock Small-scale energy cascade of the solar wind turbulence.
\newblock {\em The Astrophysical Journal}, 674(2):1153, 2008.

\bibitem{Alexandrova13}
O.~{Alexandrova}, C.~H.~K. {Chen}, L.~{Sorriso-Valvo}, T.~S. {Horbury}, and
  S.~D. {Bale}.
\newblock {Solar Wind Turbulence and the Role of Ion Instabilities}.
\newblock {\em Space Sci. Rev.}, 178(2-4):101--139, Oct 2013.

\bibitem{alexandrova2009universality}
O.~Alexandrova, J.~Saur, C.~Lacombe, A.~Mangeney, J.~Mitchell, S.~J. Schwartz,
  and P.~Robert.
\newblock Universality of solar-wind turbulent spectrum from mhd to electron
  scales.
\newblock {\em Physical review letters}, 103(16):165003, 2009.

\bibitem{bandyopadhyay2018incompressive}
R.~Bandyopadhyay, A.~Chasapis, R.~Chhiber, T.~Parashar, W.~Matthaeus, M.~Shay,
  B.~Maruca, J.~Burch, T.~Moore, C.~Pollock, et~al.
\newblock Incompressive energy transfer in the earth’s magnetosheath:
  Magnetospheric multiscale observations.
\newblock {\em The Astrophysical Journal}, 866(2):106, 2018.

\bibitem{Birn01}
J.~Birn, J.~F. Drake, M.~A. Shay, B.~N. Rogers, R.~E. Denton, M.~Hesse,
  M.~Kuznetsova, Z.~W. Ma, A.~Bhattacharjee, A.~Otto, and P.~L. Pritchett.
\newblock {GEM} magnetic reconnection challenge.
\newblock {\em J. Geophys. Res.}, 106:3715, 2001.

\bibitem{biskamp2003magnetohydrodynamic}
D.~Biskamp.
\newblock {\em Magnetohydrodynamic turbulence}.
\newblock Cambridge University Press, 2003.

\bibitem{Boldyrev17}
S.~{Boldyrev} and N.~F. {Loureiro}.
\newblock {Magnetohydrodynamic Turbulence Mediated by Reconnection}.
\newblock {\em The Astrophysical Journal}, 844:125, Aug. 2017.

\bibitem{Buechner97}
J.~Buechner and J.~P. Kuska.
\newblock Numerical simulation of three dimensional reconnection due to the
  instability of collisionless current sheets.
\newblock {\em Adv. Space Res.}, 19:1817, 1997.

\bibitem{Califano18}
F.~{Califano}, S.~S. {Cerri}, M.~{Faganello}, D.~{Laveder}, and M.~W. {Kunz}.
\newblock {Electron-only magnetic reconnection in plasma turbulence}.
\newblock {\em ArXiv e-prints}, Oct. 2018.

\bibitem{Cerri17}
S.~S. {Cerri} and F.~{Califano}.
\newblock {Reconnection and small-scale fields in 2D-3V hybrid-kinetic driven
  turbulence simulations}.
\newblock {\em New Journal of Physics}, 19(2):025007, Feb. 2017.

\bibitem{ChenEA14-grl}
C.~H.~K. {Chen}, L.~{Leung}, S.~{Boldyrev}, B.~A. {Maruca}, and S.~D. {Bale}.
\newblock {Ion-scale spectral break of solar wind turbulence at high and low
  beta}.
\newblock {\em Geophys. Res. Lett.}, 41:8081--8088, Nov. 2014.

\bibitem{ChhiberEA18}
R.~{Chhiber}, A.~{Chasapis}, R.~{Bandyopadhyay}, T.~N. {Parashar}, W.~H.
  {Matthaeus}, B.~A. {Maruca}, T.~E. {Moore}, J.~L. {Burch}, R.~B. {Torbert},
  C.~T. {Russell}, O.~{Le Contel}, M.~R. {Argall}, D.~{Fischer}, L.~{Mirioni},
  R.~J. {Strangeway}, C.~J. {Pollock}, B.~L. {Giles}, and D.~J. {Gershman}.
\newblock {Higher-Order Turbulence Statistics in the Earth's Magnetosheath and
  the Solar Wind Using Magnetospheric Multiscale Observations}.
\newblock {\em Journal of Geophysical Research (Space Physics)},
  123:9941--9954, Dec. 2018.

\bibitem{Cheyenne18}
{Computational and Information Systems Laboratory}.
\newblock {Cheyenne: HPE/SGI ICE XA System (University Community Computing)},
  2017.

\bibitem{Dong2018}
C.~{Dong}, L.~{Wang}, Y.-M. {Huang}, L.~{Comisso}, and A.~{Bhattacharjee}.
\newblock {Role of the Plasmoid Instability in Magnetohydrodynamic Turbulence}.
\newblock {\em Physical Review Letters}, 121(16):165101, Oct. 2018.

\bibitem{Drake97}
J.~F. Drake, D.~Biskamp, and A.~Zeiler.
\newblock Breakup of the electron current layer during 3-d collisionless
  magnetic reconnection.
\newblock {\em Geophys. Res. Lett.}, 24:2921, 1997.

\bibitem{eastwood2009observations}
J.~Eastwood, T.~Phan, S.~Bale, and A.~Tjulin.
\newblock Observations of turbulence generated by magnetic reconnection.
\newblock {\em Physical review letters}, 102(3):035001, 2009.

\bibitem{Franci17}
L.~Franci, S.~S. Cerri, F.~Califano, S.~Landi, E.~Papini, A.~Verdini,
  L.~Matteini, F.~Jenko, and P.~Hellinger.
\newblock Magnetic reconnection as a driver for a sub-ion-scale cascade in
  plasma turbulence.
\newblock {\em The Astrophysical Journal Letters}, 850(1):L16, 2017.

\bibitem{frisch1975possibility}
U.~Frisch, A.~Pouquet, J.~L{\'e}orat, and A.~Mazure.
\newblock Possibility of an inverse cascade of magnetic helicity in
  magnetohydrodynamic turbulence.
\newblock {\em Journal of Fluid Mechanics}, 68(4):769--778, 1975.

\bibitem{Haggerty17}
C.~C. {Haggerty}, T.~N. {Parashar}, W.~H. {Matthaeus}, M.~A. {Shay}, Y.~{Yang},
  M.~{Wan}, P.~{Wu}, and S.~{Servidio}.
\newblock {Exploring the statistics of magnetic reconnection X-points in
  kinetic particle-in-cell turbulence}.
\newblock {\em Physics of Plasmas}, 24(10):102308, Oct. 2017.

\bibitem{Haynes14}
C.~T. {Haynes}, D.~{Burgess}, and E.~{Camporeale}.
\newblock {Reconnection and Electron Temperature Anisotropy in Sub-proton Scale
  Plasma Turbulence}.
\newblock {\em The Astrophysical Journal}, 783:38, Mar. 2014.

\bibitem{hellinger2018karman}
P.~Hellinger, A.~Verdini, S.~Landi, L.~Franci, and L.~Matteini.
\newblock von k{\'a}rm{\'a}n--howarth equation for hall magnetohydrodynamics:
  Hybrid simulations.
\newblock {\em The Astrophysical Journal Letters}, 857(2):L19, 2018.

\bibitem{howes2008kinetic}
G.~Howes, W.~Dorland, S.~Cowley, G.~Hammett, E.~Quataert, A.~Schekochihin, and
  T.~Tatsuno.
\newblock Kinetic simulations of magnetized turbulence in astrophysical
  plasmas.
\newblock {\em Physical Review Letters}, 100(6):065004, 2008.

\bibitem{HowesEA08}
G.~G. Howes, S.~C. Cowley, W.~Dorland, G.~W. Hammett, E.~Quataert, and A.~A.
  Schekochihin.
\newblock A model of turbulence in magnetized plasmas: {Implications} for the
  dissipation range in the solar wind.
\newblock {\em J. Geophys. Res.}, 113, 2008.

\bibitem{Karimabadi13}
H.~{Karimabadi}, V.~{Roytershteyn}, M.~{Wan}, W.~H. {Matthaeus}, W.~{Daughton},
  P.~{Wu}, M.~{Shay}, B.~{Loring}, J.~{Borovsky}, E.~{Leonardis}, S.~C.
  {Chapman}, and T.~K.~M. {Nakamura}.
\newblock {Coherent structures, intermittent turbulence, and dissipation in
  high-temperature plasmas}.
\newblock {\em Phys. Plasmas}, 20(1):012303, Jan. 2013.

\bibitem{kolmogorov1991local}
A.~N. Kolmogorov.
\newblock The local structure of turbulence in incompressible viscous fluid for
  very large reynolds numbers.
\newblock {\em Proceedings of the Royal Society of London. Series A:
  Mathematical and Physical Sciences}, 434(1890):9--13, 1991.

\bibitem{kraichnan1967inertial}
R.~H. Kraichnan.
\newblock Inertial ranges in two-dimensional turbulence.
\newblock {\em The Physics of Fluids}, 10(7):1417--1423, 1967.

\bibitem{lazarian1999reconnection}
A.~Lazarian and E.~T. Vishniac.
\newblock Reconnection in a weakly stochastic field.
\newblock {\em The Astrophysical Journal}, 517(2):700, 1999.

\bibitem{leamon1998observational}
R.~J. Leamon, C.~W. Smith, N.~F. Ness, W.~H. Matthaeus, and H.~K. Wong.
\newblock Observational constraints on the dynamics of the interplanetary
  magnetic field dissipation range.
\newblock {\em Journal of Geophysical Research: Space Physics},
  103(A3):4775--4787, 1998.

\bibitem{Loureiro2007}
N.~F. {Loureiro}, A.~A. {Schekochihin}, and S.~C. {Cowley}.
\newblock {Instability of current sheets and formation of plasmoid chains}.
\newblock {\em Physics of Plasmas}, 14(10):100703--100703, Oct 2007.

\bibitem{lu2019turbulence}
S.~Lu, V.~Angelopoulos, A.~Artemyev, P.~Pritchett, J.~Liu, A.~Runov,
  A.~Tenerani, C.~Shi, and M.~Velli.
\newblock Turbulence and particle acceleration in collisionless magnetic
  reconnection: Effects of temperature inhomogeneity across pre-reconnection
  current sheet.
\newblock {\em The Astrophysical Journal}, 878(2):109, 2019.

\bibitem{Mallet17}
A.~{Mallet}, A.~A. {Schekochihin}, and B.~D.~G. {Chandran}.
\newblock {Disruption of sheet-like structures in Alfv{\'e}nic turbulence by
  magnetic reconnection}.
\newblock {\em Monthly Notices of the Royal Astronomical Society},
  468:4862--4871, July 2017.

\bibitem{Mandt94}
M.~E. Mandt, R.~E. Denton, and J.~F. Drake.
\newblock Transition to whistler mediated magnetic reconnection.
\newblock {\em Geophys. Res. Lett.}, 21:73, 1994.

\bibitem{matthaeus2007spectral}
W.~Matthaeus, J.~Bieber, D.~Ruffolo, P.~Chuychai, and J.~Minnie.
\newblock Spectral properties and length scales of two-dimensional magnetic
  field models.
\newblock {\em The Astrophysical Journal}, 667(2):956, 2007.

\bibitem{matthaeus1985rapid}
W.~Matthaeus and S.~Lamkin.
\newblock Rapid magnetic reconnection caused by finite amplitude fluctuations.
\newblock {\em The Physics of fluids}, 28(1):303--307, 1985.

\bibitem{matthaeus1986turbulent}
W.~Matthaeus and S.~L. Lamkin.
\newblock Turbulent magnetic reconnection.
\newblock {\em The Physics of fluids}, 29(8):2513--2534, 1986.

\bibitem{munoz2018kinetic}
P.~Mu{\~n}oz and J.~B{\"u}chner.
\newblock Kinetic turbulence in fast three-dimensional collisionless
  guide-field magnetic reconnection.
\newblock {\em Physical Review E}, 98(4):043205, 2018.

\bibitem{Osman14}
K.~T. {Osman}, W.~H. {Matthaeus}, J.~T. {Gosling}, A.~{Greco}, S.~{Servidio},
  B.~{Hnat}, S.~C. {Chapman}, and T.~D. {Phan}.
\newblock {Magnetic Reconnection and Intermittent Turbulence in the Solar
  Wind}.
\newblock {\em Phys. Rev. Lett.}, 112(21):215002, May 2014.

\bibitem{Papini18}
E.~{Papini}, L.~{Franci}, S.~{Landi}, A.~{Verdini}, L.~{Matteini}, and
  P.~{Hellinger}.
\newblock {Can Hall Magnetohydrodynamics explain plasma turbulence at sub-ion
  scales?}
\newblock {\em arXiv e-prints}, page arXiv:1810.02210, Oct. 2018.

\bibitem{parashar2009kinetic}
T.~Parashar, M.~Shay, P.~Cassak, and W.~Matthaeus.
\newblock Kinetic dissipation and anisotropic heating in a turbulent
  collisionless plasma.
\newblock {\em Physics of Plasmas}, 16(3):032310, 2009.

\bibitem{Phan07}
T.~D. Phan, J.~F. Drake, M.~A. Shay, F.~S. Mozer, and J.~P. Eastwood.
\newblock Evidence for an elongated ($>60$ ion skin depths) electron diffusion
  region during fast magnetic reconnection.
\newblock {\em Phys. Rev. Lett.}, 99:255002, 2007.

\bibitem{Phan18}
T.~D. {Phan}, J.~P. {Eastwood}, M.~A. {Shay}, J.~F. {Drake}, B.~U.~{\"O}.
  {Sonnerup}, M.~{Fujimoto}, P.~A. {Cassak}, M.~{{\O}ieroset}, J.~L. {Burch},
  R.~B. {Torbert}, A.~C. {Rager}, J.~C. {Dorelli}, D.~J. {Gershman},
  C.~{Pollock}, P.~S. {Pyakurel}, C.~C. {Haggerty}, Y.~{Khotyaintsev},
  B.~{Lavraud}, Y.~{Saito}, M.~{Oka}, R.~E. {Ergun}, A.~{Retino}, O.~{Le
  Contel}, M.~R. {Argall}, B.~L. {Giles}, T.~E. {Moore}, F.~D. {Wilder}, R.~J.
  {Strangeway}, C.~T. {Russell}, P.~A. {Lindqvist}, and W.~{Magnes}.
\newblock {Electron magnetic reconnection without ion coupling in Earth's
  turbulent magnetosheath}.
\newblock {\em Nature}, 557:202--206, May 2018.

\bibitem{politano1998karman}
H.~Politano and A.~Pouquet.
\newblock von k{\'a}rm{\'a}n--howarth equation for magnetohydrodynamics and its
  consequences on third-order longitudinal structure and correlation functions.
\newblock {\em Physical Review E}, 57(1):R21, 1998.

\bibitem{pucci2018generation}
F.~Pucci, W.~H. Matthaeus, A.~Chasapis, S.~Servidio, L.~Sorriso-Valvo,
  V.~Olshevsky, D.~Newman, M.~Goldman, and G.~Lapenta.
\newblock Generation of turbulence in colliding reconnection jets.
\newblock {\em The Astrophysical Journal}, 867(1):10, 2018.

\bibitem{Retino07}
A.~{Retin{\`o}}, D.~{Sundkvist}, A.~{Vaivads}, F.~{Mozer}, M.~{Andr{\'e}}, and
  C.~J. {Owen}.
\newblock {In situ evidence of magnetic reconnection in turbulent plasma}.
\newblock {\em Nature Physics}, 3:236--238, Apr. 2007.

\bibitem{sahraoui2009evidence}
F.~Sahraoui, M.~Goldstein, P.~Robert, and Y.~V. Khotyaintsev.
\newblock Evidence of a cascade and dissipation of solar-wind turbulence at the
  electron gyroscale.
\newblock {\em Physical review letters}, 102(23):231102, 2009.

\bibitem{servidio2009magnetic}
S.~Servidio, W.~Matthaeus, M.~Shay, P.~Cassak, and P.~Dmitruk.
\newblock Magnetic reconnection in two-dimensional magnetohydrodynamic
  turbulence.
\newblock {\em Physical review letters}, 102(11):115003, 2009.

\bibitem{servidio2010statistics}
S.~Servidio, W.~Matthaeus, M.~Shay, P.~Dmitruk, P.~Cassak, and M.~Wan.
\newblock Statistics of magnetic reconnection in two-dimensional
  magnetohydrodynamic turbulence.
\newblock {\em Physics of Plasmas}, 17(3):032315, 2010.

\bibitem{sharma2019transition}
P.~Sharma~Pyakurel, M.~Shay, T.~Phan, W.~Matthaeus, J.~Drake, J.~TenBarge,
  C.~Haggerty, K.~Klein, P.~Cassak, T.~Parashar, et~al.
\newblock Transition from ion-coupled to electron-only reconnection: Basic
  physics and implications for plasma turbulence.
\newblock {\em Physics of Plasmas}, 26(8):082307, 2019.

\bibitem{Shay98a}
M.~A. Shay, J.~F. Drake, R.~E. Denton, and D.~Biskamp.
\newblock Structure of the dissipation region during collisionless magnetic
  reconnection.
\newblock {\em J. Geophys. Res.}, 103:9165, 1998.

\bibitem{Shay18}
M.~A. {Shay}, C.~C. {Haggerty}, W.~H. {Matthaeus}, T.~N. {Parashar}, M.~{Wan},
  and P.~{Wu}.
\newblock {Turbulent heating due to magnetic reconnection}.
\newblock {\em Physics of Plasmas}, 25(1):012304, Jan. 2018.

\bibitem{shebalin1983anisotropy}
J.~V. Shebalin, W.~H. Matthaeus, and D.~Montgomery.
\newblock Anisotropy in mhd turbulence due to a mean magnetic field.
\newblock {\em Journal of Plasma Physics}, 29(3):525--547, 1983.

\bibitem{smith2006dependence}
C.~W. Smith, K.~Hamilton, B.~J. Vasquez, and R.~J. Leamon.
\newblock Dependence of the dissipation range spectrum of interplanetary
  magnetic fluctuationson the rate of energy cascade.
\newblock {\em The Astrophysical Journal Letters}, 645(1):L85, 2006.

\bibitem{Sonnerup79}
B.~U.~{\"O.}. Sonnerup.
\newblock Magnetic field reconnection.
\newblock In L.~J. Lanzerotti, C.~F. Kennel, and E.~N. Parker, editors, {\em
  Solar System Plasma Physics}, volume~3, page~46. North Halland Pub.,
  Amsterdam, 1979.

\bibitem{stawarz2019properties}
J.~Stawarz, J.~Eastwood, T.~Phan, I.~Gingell, M.~Shay, J.~Burch, R.~Ergun,
  B.~Giles, D.~Gershman, O.~Le~Contel, et~al.
\newblock Properties of the turbulence associated with electron-only magnetic
  reconnection in earth’s magnetosheath.
\newblock {\em The Astrophysical Journal Letters}, 877(2):L37, 2019.

\bibitem{strauss1988turbulent}
H.~Strauss.
\newblock Turbulent reconnection.
\newblock {\em The Astrophysical Journal}, 326:412--417, 1988.

\bibitem{verma2005energy}
M.~K. Verma, A.~Ayyer, and A.~V. Chandra.
\newblock Energy transfers and locality in magnetohydrodynamic turbulence.
\newblock {\em Physics of plasmas}, 12(8):082307, 2005.

\bibitem{wilder2017multipoint}
F.~Wilder, R.~Ergun, S.~Eriksson, T.~Phan, J.~Burch, N.~Ahmadi, K.~Goodrich,
  D.~Newman, K.~Trattner, R.~Torbert, et~al.
\newblock Multipoint measurements of the electron jet of symmetric magnetic
  reconnection with a moderate guide field.
\newblock {\em Physical review letters}, 118(26):265101, 2017.

\bibitem{Yamada10}
M.~{Yamada}, R.~{Kulsrud}, and H.~{Ji}.
\newblock {Magnetic reconnection}.
\newblock {\em Rev. Modern Phys.}, 82:603, Jan. 2010.

\bibitem{zeiler2002three}
A.~Zeiler, D.~Biskamp, J.~Drake, B.~Rogers, M.~Shay, and M.~Scholer.
\newblock Three-dimensional particle simulations of collisionless magnetic
  reconnection.
\newblock {\em Journal of Geophysical Research: Space Physics}, 107(A9):SMP--6,
  2002.

\end{thebibliography}

\end{document}